\documentclass{article}
\PassOptionsToPackage{numbers, compress}{natbib}

    \usepackage[preprint]{neurips_2024}

\usepackage[utf8]{inputenc} 
\usepackage[T1]{fontenc}    
\usepackage{hyperref}       
\usepackage{url}            
\usepackage{booktabs}       
\usepackage{amsfonts}       
\usepackage{nicefrac}       
\usepackage{microtype}      
\usepackage{xcolor}         
\usepackage{comment}        
\usepackage{graphicx}       
\usepackage{amsmath}
\usepackage{authblk}

\title{Scaling Law in Neural Data: Non-Invasive Speech Decoding with 175 Hours of EEG Data}
\author[1]{Motoshige Sato}
\author[1]{Kenichi Tomeoka}
\author[1]{Ilya Horiguchi}
\author[1]{Kai Arulkumaran}
\author[1]{Ryota Kanai}
\author[1,*]{Shuntaro Sasai}
\affil[1]{Araya Inc. Tokyo, Japan\\}
\affil[*]{Correspondence: Shuntaro Sasai, Ph.D. (sasai\_shuntaro@araya.org)}
\begin{document}

\maketitle

\begin{abstract}
Brain-computer interfaces (BCIs) hold great potential for aiding individuals with speech impairments. Utilizing electroencephalography (EEG) to decode speech is particularly promising due to its non-invasive nature. However, recordings are typically short, and the high variability in EEG data has led researchers to focus on classification tasks with a few dozen classes. To assess its practical applicability for speech neuroprostheses, we investigate the relationship between the size of EEG data and decoding accuracy in the open vocabulary setting. We collected extensive EEG data from a single participant (175 hours) and conducted zero-shot speech segment classification using self-supervised representation learning. The model trained on the entire dataset achieved a top-1 accuracy of 48\% and a top-10 accuracy of 76\%, while mitigating the effects of myopotential artifacts. Conversely, when the data was limited to the typical amount used in practice ($\sim$10 hours), the top-1 accuracy dropped to 2.5\%, revealing a significant scaling effect. Additionally, as the amount of training data increased, the EEG latent representation progressively exhibited clearer temporal structures of spoken phrases. This indicates that the decoder can recognize speech segments in a data-driven manner without explicit measurements of word recognition. This research marks a significant step towards the practical realization of EEG-based speech BCIs.

\end{abstract}

\section{Introduction}
Motor speech disorder is a severe medical condition that leaves people barely or completely unable to speak, and occurs in 90\% of Parkinson's disease patients \citep{moya2019parkinson}, 45.2\% of stroke patients \citep{bahia2016dysarthria}, and 95\% of amyotrophic lateral sclerosis (ALS) patients \citep{ball2004communication}. Typical communication aids for speech impairments, such as those using eye trackers, are significantly slower than spontaneous speech, especially in the late stages of ALS, where vision loss and eye movement deficits can cause fatigue problems and render use impossible \citep{hayashi2003patients,san2010evaluation,ball2010eye,spataro2014eye,chen2018eye}. Conversely, among recent advances in speech brain-computer interfaces (BCIs), invasive neural recordings such as intracortical microelectrode arrays and cortical electrograms (ECoG) have shown remarkable promise in achieving word production rates close to those of natural speech \citep{anumanchipalli2019speech,makin2020machine,willett2021high,moses2021neuroprosthesis,willett2023high,metzger2023high}, while also providing a less burdensome alternative for users. However, these methods, which require surgery to implant electrodes in the brain, are highly invasive and present significant psychological and physical barriers. For this reason, there is a growing demand for a speech BCI using non-invasive neural recording technology, which has far lower barriers for use.

Among non-invasive recordings of neural activity, EEG enables portable and real-time BCI. fMRI and MEG require large magnetic field equipment and hence are impractical for everyday use. NIRS, on the other hand, is a neural recording method that is suitable for everyday use. However, because it measures blood flow changes as a secondary effect of neural activity, its temporal resolution is insufficient for real-time decoding of continuous speech, where syllables occur at intervals of a few hundred milliseconds. EEG is a non-invasive and routinely available recording technique that can capture features of speech with a high enough temporal resolution at a relatively low cost.

On the other hand, EEGs measure signals that pass through the skull and skin, which weaken the signal amplitudes and are prone to noise and myopotential artifacts. Furthermore, since neural activity exhibits significant day-by-day variability \citep{huang2023discrepancy}, EEGs of the same subject are rarely recorded for long periods of time across days except for epilepsy monitoring or clinical purposes \citep{renzel2021sensitivity}. In fact, speech BCI systems utilizing EEG have collected data for each individual for only a few hours at most. Moreover, since the analysis has previously been limited to a maximum lexicon size of 13 words \citep{lee2020eeg,lee2023towards,koizumi2018development,watanabe2020synchronization,sereshkeh2017eeg,sereshkeh2017online}, it has not yet achieved a practical level that allows for highly flexible communication. To go beyond this and decode under open vocabulary conditions without limiting the number of words would therefore require a large amount of speech data containing a variety of words and phrases. Even then, it is not clear whether collecting such data by recording across many days would improve decoding performance because of the large day-by-day variability of EEG. In this work we investigated this problem, assessing decoding performance by recording the EEG of one subject during speech, over several months, for a total of 175 hours.

We employed the contrastive language-image pre-training (CLIP) \citep{radford2021learning} method, which uses self-supervised learning and hence does not require labels, on a large amount of paired EEG and speech data. Using CLIP we integrate the different modalities of EEG and speech and acquire representations that can be easily used for various downstream tasks, such as zero-shot classification of phrases, and speech reconstruction.

\begin{figure} [h] 
    \label{fig:diagram}
    \centering
    \includegraphics[width=1.0\linewidth]{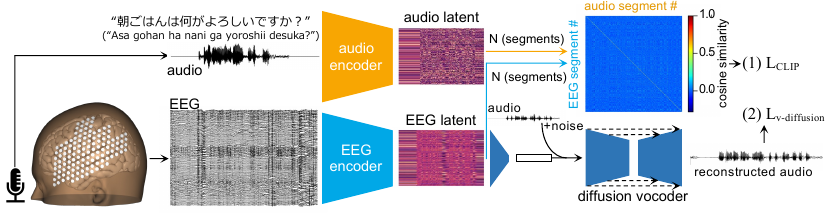}
    \caption{Decoding framework. EEG and speech were recorded simultaneously and converted to latent representations by a (fixed) audio encoder and an EEG encoder, respectively, for each 5-second segment. In step (1), the CLIP loss is applied on the N-segment pairwise cosine similarity matrix. In step (2), a diffusion vocoder was trained to reconstruct the speech waveform from the EEG latent representation.}
\end{figure}

\section{Related Work}
Here we review studies that have attempted to decode words from EEG in an open vocabulary setting.

Défossez et.al. \citep{defossez2023decoding} evaluated the zero-shot classification accuracy of EEG segments while participants listened to speech by optimizing the CLIP loss between the latent representations of EEG and audio. They used two datasets consisting of 19.2 h EEG from 19 participants, and 6.7 h EEG from 33 participants \citep{Broderick2018eegdataset, Brennan2019eegdataset}. They succeeded in achieving 25.7\% in top-10 accuracy out of 190 segments for the former, and 17.7\% out of 1,842 segments for the latter. The decoding performance was 5-30 times higher than chance levels. However, it is important to note that their experimental paradigm focused on decoding speech perception rather than directly targeting the decoding of the participant's own verbal output.

 \citep{wang2022open,duan2023dewave} conducted decoding EEG during silent text comprehension, leveraging an EEG dataset known as ZuCo \citep{hollenstein2018zuco,hollenstein2019zuco}, in which data was collected from 18 participants in total for 100-180 minutes per subject. Their approach involved the integration of eye-tracking data concurrent with EEG recordings to decode the process of reading. The EEG-to-text translation was conducted by aligning the latent representation of individual words with corresponding EEG signals captured during eye fixation periods, utilizing a machine-translation-based method \citep{wang2022open} and a CLIP-based loss function with vector quantization \citep{duan2023dewave}. However, subsequent to the publication of their findings, the researchers acknowledged an overestimate in their accuracy due to the usage of teacher-forcing during evaluation. When teacher-forcing was omitted, they were unable to achieve performance significantly above random \citep{jo2024are}. Liu et al. \citep{liu2024eeg2text} successfully improved BLEU and ROUGE scores by up to 5\% compare to DeWave\citep{duan2023dewave} by implementing a multi-view transformer that combines a brain area independent transformer with an integrated transformer after pre-training with a masked auto-encoder and on the same ZuCo dataset.

\section{Methods}

\subsection{Dataset} 
\begin{table} [h]
    \caption{Dataset properties. We summarized the properties of the two EEG datasets \citep{Broderick2018eegdataset,Brennan2019eegdataset} used by Défossez et.al. \citep{defossez2023decoding}, and our dataset. Word overlap indicates the proportion of unique words in the test dataset that also appear in the training dataset. Japanese word segmentation on our dataset was performed using the janome tokenizer \citep{janome}.}
    \label{tab:dataset properties}
    \centering
    \begin{tabular}{ccccccccc}
      \toprule
      & & & & & & \multicolumn{2}{c}{vocabulary size} &                                             \\
      \cmidrule(r){7-8}
      dataset & language & task & channels & subjects & duration & train & test & word overlap \\
      \midrule
      \citep{Brennan2019eegdataset} & EN & listening & 60  & 33 & 6.7 [h]  & 513  & 148 & 60\% \\
      \citep{Broderick2018eegdataset}
      & EN & listening & 128 & 19 & 19.2 [h] & 1418 & 764 & 67\%                           \\
      ours  & JP & speech  & 128 & 1  & 175 [h]  & 28159  & 10103 & 85\%                           \\
      \bottomrule
    \end{tabular}    
\end{table}

\subsubsection{Participant} \label{sec:participant}
A healthy male adult participated in this study. He had no history of neurological or psychiatric illnesses. He was initially given a briefing on the purpose of the study and experimental protocol, and he provided informed consent before we proceeded to the actual experiment. Our study obtained ethical approval by the Shiba Palace Clinic Ethics Review Committee. The experiments were undertaken in compliance with national legislation and the Code of Ethical Principles for Medical Research Involving Human participants of the World Medical Association (the Declaration of Helsinki).

\subsubsection{EEG and EMG recording} \label{sec:recording}
EEG, electrooculogram (EOG), upper and lower orbicularis oris electromyogram (EMG), and speech voice signals were recorded simultaneously while the participant read aloud dialogues from text corpora \citep{sonobe2017jsut}, novels, and text-based games displayed on a computer screen. During the recording, pace instruction cues were not presented, and the subject continuously read the text aloud at a natural speed. The data recording experiment was conducted over a period of 48 days, with a total recording duration of 175 hours.
EEG electrode placement was targeted around the language areas of the left hemisphere (Fig \ref{fig:diagram}) and eight g.pangolin \citep{lee2022individual,schreiner2024mapping} electrode sheets (16 electrodes per sheet, g.tec, Austria) were placed over Broca's area, auditory area, Wernicke's area, and premotor area.
For EOG recording, the electrodes were placed above and below the left eye. For measurement of mouth muscle activity, each of two electrodes were placed on each of the left upper and lower orbicularis oris.

\subsubsection{Preprocessing} \label{sec:preprocess}
EEG was acquired in real-time at 1200 Hz using the g.NEEDaccess Python API (g.tec, Austria). We denoised the raw EEG using MNE-Python \citep{GramfortEtAl2013mne}; details are shown in Figure \ref{fig:architecture}a. To reduce the EMG contamination in the EEG signal, we used an adaptive filter \citep{diniz1997adaptive,haykin2002adaptive}, through which the EEG signal linearly predicted from the EMG signals (EOG, and upper and lower orbicularis oris EMG) was removed from the EEG. The adaptive filter has been used in several studies \citep{correa2007artifact,kher2016adaptive,molla2012artifact} as a technique to remove EMG artifacts from EEG as it has a fast computation time and hence is suitable for real-time BCI. Details of the adaptive filter are described in Appendix \ref{sec:adaptive fitler}.

The participant's spoken voice was recorded with a microphone at 48 kHz and downsampled to 16 kHz. Furthermore, we applied Silero VAD \citep{SileroVAD} to remove noise from the audio signal. After synchronizing the timestamps of the EEG and audio data, the signals were epoched into 5-second time windows without overlap, ensuring temporal correspondence between the EEG and audio modalities. For each window, the EEG data were Z-scored along the time axis and clipped to a range of ±5. Windows containing less than 20\% of speech segments, as detected by Silero VAD, were excluded.

\subsubsection{Data splits} \label{sec:data split}
  EEG data, like other time series data, is known to exhibit temporal trends and variations. Therefore, in order to test the ability for our decoder to generalise over time, the dataset was split with chronological order, with the initial 80\% assigned to train set, the following 10\% to validation set, and the final 10\% to test set.

\subsection{Model}

\subsubsection{EEG preprocessing and encoder} \label{sec:decoding architecture}
\begin{figure} [h]
    \centering
        \includegraphics[width=0.8\linewidth]{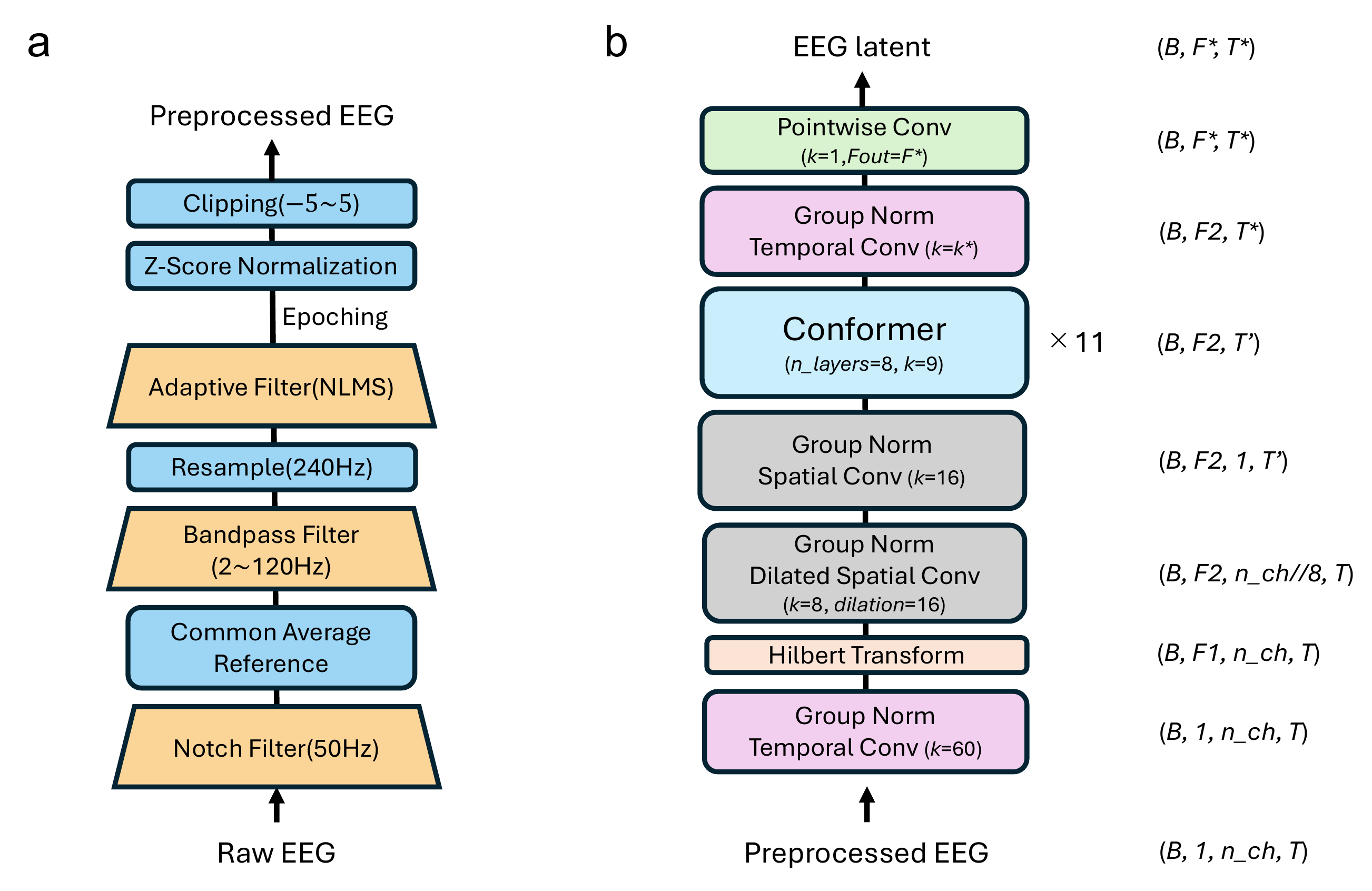}
        \caption{A diagram of EEG preprocessing and EEG encoder. (a) The pre-processing procedure. (b) EEG encoder architecture. The number of feature dimensions $F^*$ and the number of time steps $T^*$ in the latent representation differ depending on the audio encoder. The output shape of each layer is shown in the \textit{right} column.}
        \label{fig:architecture}
\end{figure}
For the EEG encoder (Figure \ref{fig:architecture}), we developed a model combining HTNet \citep{lawhern2018eegnet,peterson2021generalized} and Conformer \citep{Gulati2020conformer}. For the audio encoder, we used the following three pre-trained models: wav2vec2.0 \citep{Baevski2020wav2vec2}, Whisper \citep{Radford2022whisper} (only the encoder module was used), and Encodec \citep{Defossez2022encodec}.

For speech reconstruction, a 1D U-Net was trained to convert EEG latents to speech waveforms using diffusion vocoder architecture \citep{schneider2023mousai}.

\subsubsection{Decoder training} \label{sec:decoder training}
Encoders were applied to the simultaneously recorded speech and EEG signals, respectively, and the CLIP loss between EEG and audio latent representations was optimized. The loss function is expressed in Equation \ref{eq:CLIP}:
\begin{equation}
\begin{split}
L_{\text{CLIP}} &= \text{CrossEntropy}(I,\hat{X}\hat{Y} ) \\
               &= -\frac{1}{N} \sum_{i}^{N} \log \left( \frac{\exp(\hat{X}_i \cdot \hat{Y}_i )}{\sum_{j}^{N} \exp(\hat{X}_i \cdot \hat{Y}_j )} \right)\label{eq:CLIP}
\end{split}
\end{equation}

where, $I$ denotes the identity matrix, $N$ denotes the batch size, and $x$ and $y$ denote the audio and EEG data, respectively. $X = E_{\text{EEG}}(x)$ and, $Y = E_{\text{audio}}(y)$, where $E_{\text{EEG}}$ and $E_{\text{audio}}$ denote the EEG encoder and audio encoder, respectively. The normalized representations are obtained as $\hat{X}_i = X_i / \|X_i\|$ and $\hat{Y}_i = Y_i / \|Y_i\|$.

For the audio encoder, a pre-trained wav2vec2.0, Whisper or Encodec model was applied, with weights kept fixed. When performing CLIP training of the EEG encoder with wav2vec2.0 embeddings, the randomly initialized EEG encoder was trained for 300 epochs with a batch size of 512 samples and the Lamb optimizer \citep{You2019lamb} (initial learning rate: 0.001, weight decay: 0.01). We used a cosine annealing learning rate schedule, with 1000 iterations of warmup (corresponding to 7.8 epochs). For CLIP training with either the Encodec or Whisper audio encoders, we instead used AdamW \citep{Loshchilov2017adamw} (learning rate: 0.0001, weight decay: 0.01),since it achieved better performance than Lamb in a pilot study. The learning rate scheduler was the same as with wav2vec2.0.

For speech reconstruction, the diffusion vocoder was trained for 300 epochs with a batch size of 512 samples and the AdamW optimizer (learning rate: 0.001, weight decay: 0.01). The reconstructed speech waveform was generated efficiently by distillation model \citep{salimans2022progressive} of the vocoder with DDIM sampling \citep{song2021denoising}. 1000 time steps were used for the denoising process.

\subsection{Evaluation}
Generalization performance was evaluated by performing inference on the test dataset using the weights from the epoch with the lowest loss on the validation dataset.

To assess the effectiveness of pretraining using CLIP, zero-shot segment classification was conducted, following the same procedure as Défossez et.al. \citep{defossez2023decoding}. The cosine similarity matrix was computed between the EEG latents and audio latents of the 512 samples in the test set. For each EEG latent, the indices of the top-k most similar audio latents based on cosine similarity were extracted. The top-k accuracy was defined as the percentage of EEG latents for which the corresponding audio latent was among the top-k most similar latents. In other words, this score is higher when the latent predicted from the EEG is more similar to the corresponding audio latent than to the latents of other samples. The average performance ± 1 standard deviation (SD) was obtained from 16 batches consisting of 512 test samples per batch for Table \ref{tab:CLIP_accuracy}, \ref{tab:latent VAD}, \ref{tab:emg-table} and Figure \ref{fig:data scaling}, \ref{fig:latent VAD}.

For the evaluation of speech reconstruction, we computed the mel-cepstral distortion (MCD) metric, as used in previous studies \citep{anumanchipalli2019speech,metzger2023high}. MCD is a widely accepted objective measure of the dissimilarity between two speech signals in the mel-cepstral domain \citep{kubichek1993mel}, and is formulated by the following equation \ref{eq:mcd}:

\begin{equation}
\begin{split}
\text{MCD}(\hat{y},y)=\frac{10}{\log (10)}\sqrt{(\mathop{\sum }\limits_{d=1}^{24}{({\text{mc}}_{d}^{y}-{\text{mc}}_{d}^{\hat{y}})}^{2})},\label{eq:mcd}
\end{split}
\end{equation}

where, $\hat{y}$ and $y$ denotes the reconstructed speech and the recorded speech respectively, and ${\text{mc}}_{d}$ denotes the $d$th dimension of the mel-cepstrum. A lower MCD value indicates a higher degree of similarity between the reconstructed and original speech signals. The average performance ± 1 SD was obtained from 8448 test samples for Figure \ref{fig:reconstruction}.

\section{Results}

\subsection{Zero-shot speech segment classification} \label{tab:zeroshot accuracy}
\begin{table} [h]
    \caption{Zero-shot segment classification accuracy across different audio encoders.}
    \label{tab:CLIP_accuracy}
    \centering
    \begin{tabular}{l||ll}
      \toprule
      Audio Encoder & top1  & top10 (\%)            \\
      \midrule
      wav2vec2.0    & \textbf{48.5} & \textbf{76.0} \\
      Whisper       & 10.6  & 52.9                  \\
      Encodec       & 28.8  & 60.4                  \\
      \bottomrule
    \end{tabular}    
\end{table}
Table \ref{tab:CLIP_accuracy} presents the average zero-shot speech segment classification accuracy for 512 samples of decoders trained with all training data. The EEG encoder was trained to optimize the CLIP loss with the respective audio latent of the pre-trained audio encoders: wav2vec2.0, Encodec and Whisper. Among the three types of speech encoders, wav2vec2.0 achieved the highest accuracy, with 48.5\% for top-1 accuracy and 76.0\% for top-10 accuracy. This performance is significantly higher than the chance level (top-1: 0.19\%, top-10: 1.95\%).

\subsection{Data scaling in classification accuracy} \label{sec:data scaling}
\begin{figure} [h]
\centering
    \includegraphics[width=1.0\linewidth]{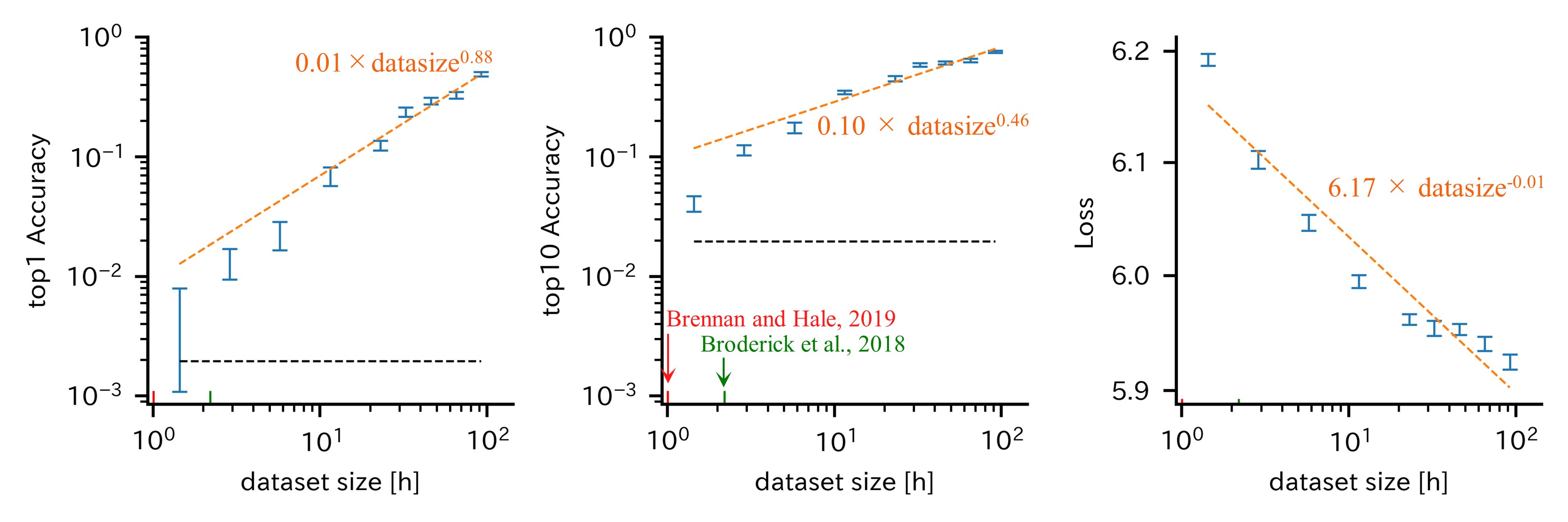}
    \caption{Data scaling. The relationship between the training dataset size (total segment length) and the top-1 accuracy (\textit{left}), top-10 accuracy(\textit{center}), and loss (\textit{right}) on the test dataset. The \textit{black} dashed line indicates chance level, and the \textit{orange} dashed line indicates the best linear fit to the data. The dataset sizes of the \textit{green} and \textit{red} arrows are the dataset sizes of the datasets \citep{Broderick2018eegdataset,Brennan2019eegdataset} used in  \citep{defossez2023decoding}, respectively.}
    \label{fig:data scaling}
    
\end{figure}

Figure \ref{fig:data scaling} illustrates the relationship between zero-shot segment classification accuracy and the EEG encoder training loss versus the amount of training data. A common test set was applied to all models of each training data amount to evaluate performance. Figure \ref{fig:data scaling}(\textit{left}, \textit{center}) show that the classification accuracy improves as the amount of training data increases. Furthermore, the accuracy has not saturated with respect to the training data and continues to rise, suggesting that increasing the amount of data further would lead to further improvements in accuracy.

\label{sec:latent VAD}
\begin{figure} [h]

    \centering
    \includegraphics[width=1.0\linewidth]{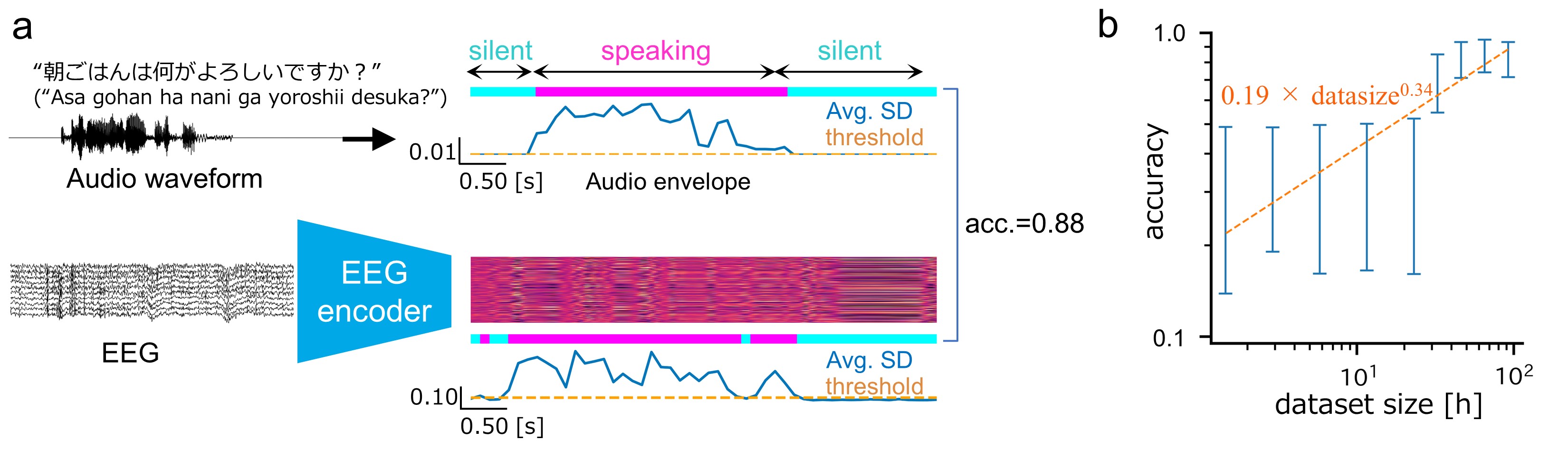}
    \caption{Voice activity detected from EEG latent representations without explicit training. (a) Process of speech interval detection. The speech waveform (\textit{upper left}) and EEG (\textit{lower left}) were each converted into a latent representation (\textit{upper right color map}) through the encoder, and the variance for each feature dimension was taken in a sliding window of 100 ms and then averaged across feature domain (\textit{lower blue line}). Intervals above the threshold (\textit{orange line}) for this value were detected as speaking periods, and intervals below the threshold were detected as silent periods. The speech segment for the ground truth was determined by applying a threshold value to the waveform envelop. In this example, the overlap between the speech segments and the segments detected by the EEG latent (accuracy) was 0.88. (b) The relationship between the speech detection accuracy and the training dataset size.
    }
    \label{fig:latent VAD}
\end{figure}

\begin{table} [h]
    \caption{Comparison of voice activity detection performance between different latents and raw EEG.}
    \label{tab:latent VAD}
    \centering
    \begin{tabular}{c||ccc}
      \toprule
      modality &predictor      & accuracy             & balanced accuracy (\%) \\
      \midrule
      audio &wav2vec2.0 latent & \textbf{99.9 ± 0.45} & \textbf{99.6 ± 3.3}    \\
      EEG &EEG latent          & 88.6 ± 10.3          & 78.7 ± 16.0            \\
      EEG &raw EEG             & 76.8 ± 22.4          & 60.6 ± 20.4            \\
      \bottomrule
    \end{tabular}    
\end{table}

Upon observing the EEG latent and audio latent representations, both exhibit similar temporal variations (Figure \ref{fig:diagram}). The intervals without speech (periods between syllables) show nearly constant temporal variations across each feature dimension. The EEG latent and wav2vec2.0 latent representation were divided into 50 bins (=100ms) respectively, and the SD values were calculated along the time axis within each bin. These SD values were averaged across the features and binarized using a threshold to detect the onset of each syllable. 
The ground truth for speech intervals was obtained by dividing the envelope of the speech waveform within each 5-second time window into 50 bins (=100 ms). The maximum value within each bin was taken as the representative value for that bin where the representative value exceeded a predetermined threshold were considered as speech, while those below the threshold were regarded as non-speech.
Figure \ref{fig:latent VAD}(a) illustrates the coincidence rate between the speech segments predicted from the EEG latent representation and those determined from the wav2vec2.0 latent representation. Figure \ref{fig:latent VAD}(b) also confirms the data scaling of the speech segment agreement rate with respect to the learning data.
The thresholds for determining speech intervals were set to 0.001 for the speech waveform, 0.047 for the speech latent representation, 0.01 for the raw EEG waveform, and 0.06 for the EEG latent representation. These values were obtained through a grid search to maximize the interval overlaps between the speech waveform and each respective data type.

\subsection{Voice reconstruction from EEG latent representations} \label{sec:voice reconst}
\begin{figure} [h]
    \label{fig:reconstruction}
        \centering
        \includegraphics[width=1.0\linewidth]{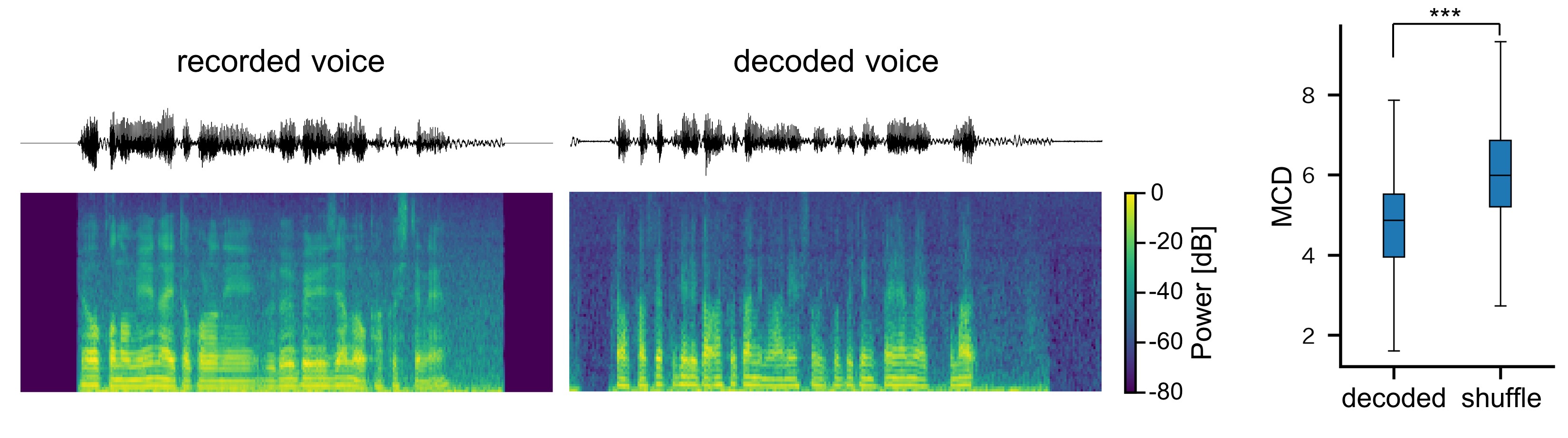}
        \caption{Representative recorded voice (\textit{left}) and reconstructed voice (\textit{middle}). The \textit{top} panels show the voice waveforms and the \textit{bottom} panels are mel-spectrograms. MCD scores (\textit{right}), where smaller values indicate better performance. These were compared to a random model trained on a dataset with shuffled EEG and speech correspondences. The box plot illustrates the distribution of the scores obtained from 8,448 test samples. Decoding performance significantly outperformed chance ($p^{***}<{10}^{-3}, {t}_{8447}=-73.5$, paired \textit{t}-test).}
        \label{fig:voice reconst}
\end{figure}

The process of reconstruction of the participant's spoken voice from EEG is shown in Figure \ref{fig:voice reconst}. After pre-training of EEG and audio latent representations with CLIP, a Vocoder diffusion model \citep{salimans2022progressive} was trained to convert EEG latent representations into audio waveforms, with fixed EEG encoder weights. The performance was evaluated by MCD, as with previous studies \citep{anumanchipalli2019speech,metzger2023high}. The performance was significantly better than the MCD of models trained on the data where EEG and audio pairs were shuffled. We achieved an MCD of 4.68 dB on average, which is comparable to the results of SOTA (3.8-5.0 dB) of invasive speech decoding \citep{anumanchipalli2019speech,metzger2023high}. It should be noted, however, that the MCD index is affected by differences in the sampling rate of the audio and silent segments, therefore a direct comparison is not feasible. While the reconstructed speech bears a resemblance to the participant's voice, it remains challenging to clearly discern the content of the speech. For the practical implementation of speech BCIs, it is still desirable to achieve a high level of clarity in the reconstructed speech. Consequently, improving the quality of the reconstructed speech remains a challenge for future research.

\subsection{Influence of EMG contamination on decoding performance}
\label{sec:emg-test}

\begin{table} [h]
    \caption{Low susceptibility to EMG artifacts. The susceptibility to EMG artifacts of EEG encoder was evaluated by the top-10 decoding accuracy from EMG. This score was evaluated for EOG, upper orbicularis oris EMG (uEMG), lower orbicularis oris EMG (lEMG), and the mean of these three EMGs (mEMG) in each column. In order to train the model for inference independent of EMG, we applied data augmentation with synthetic data, mixing EMG scaled by a factor of $\alpha$ from different trials to EEG, which was multiplied by a factor of $1-\alpha$. Here, $\alpha$ is a random value within range [0, 0.95]. The model was trained to minimize the CLIP loss with the audio latent for the EEG trial while ignoring the audio latent of the trial on the EMG side. Fig \ref{sec:emg-test-proceduer} provides a schematic overview of this operation. Each row indicates the EMG type mixed in this data synthesis.}
    \label{tab:emg-table}
    \centering
    \begin{tabular}{l||llllc}
      \toprule
      \multicolumn{5}{r}{Predictor \& top-10 accuracy(\%)} \\
      \cmidrule(r){1-6}
      augment w/  & EOG↓ & uEMG↓ & lEMG↓ & mEMG↓ & EEG↑        
      \\
      \midrule
      -- & 28.8 & 14.6 & 13.8 & 24.6  & 76.0                           \\
      EOG   & 5.02 & 6.47 & 7.04 & 6.38  & 72.3                                    \\
      uEMG  & 2.88 & 2.95 & 3.05 & 2.69 & 70.1 \\
      lEMG  & 4.47 & 4.97 & 5.49 & 6.89 & 69.0                                     \\
      mEMG  & 3.19 & 4.43 & 4.53 & 4.09 & 70.9                                     \\
      \bottomrule
    \end{tabular}
  \end{table}

EEG is easily contaminated by myopotential artifacts\citep{porcaro2015removing,jiang2019removal}. If these signals other than neural activities contribute to speech decoding, this technology will not work when used by patients with speech impairments, since similar levels of muscle activity will not be obtained from those patients. Therefore, to establish that speech can be reliably decoded from EEG independently of myopotential artifacts, we examined whether the decoder's inference performance was reduced when EMG was used as the input instead of EEG. To this end, we propose a method to train the decoder to infer audio latent representations while ignoring EMG by data augmentation where EMG signals from different trials were artificially added to EEG signals.

Table \ref{tab:emg-table} shows that without data augmentation, the inference accuracy is significantly lower than the EEG decoding accuracy, but the speech is still decoded from EMG significantly above 2\% of chance level. In other words, the accuracy of the EEG Encoder cannot be explained solely by the effect of EMG, but the influence of EMG artifacts cannot be completely denied.

The top-10 decoding performance from EMG signals dropped to around 3\% for the models trained to disregard EMG signals using the mixed EMG and EEG signals This decline of performance to the near chance level of 2\% suggests that the inference from EEG was not heavily influenced by EMG artifacts. Additionally, the decoding accuracy for EEG input remained at approximately 70\% as measured by the top-10 accuracy, demonstrating the ability to decode speech without depending on muscle-related artifacts.

\section{Limitations} \label{sec:limitation}
Our current study achieved a top-1 accuracy of 48\% on the 512-phrase classification task under open vocabulary conditions, which is unprecedented for EEG-based speech decoding. However, it is important to note that this does not imply that the system can be immediately used as a speech neuroprosthesis. Several challenges must be addressed in future work to fully develop a practical speech BCI.

Although we collected an exceptionally large sample of 175 hours of data, the data was obtained only from a single participant. As such, it is unclear whether this system can be transferred to other participants. However, we anticipate that it is possible to achieve a generally usable speech BCI without collecting such large-scale data from all participants. Typically, models pre-trained on large datasets can be fine-tuned with smaller amounts of new data. We hypothesize that this empirical observation will also hold for speech decoding. Future research should investigate the data amount required to enable transferability across participants.

Many highly performant speech neuroprostheses methods have been developed for invasive measurements from individuals with speech disabilities attempting speech. In contrast, our study demonstrates that high-accuracy speech decoding is possible with non-invasive measurements by collecting data from healthy individuals during overt speech. This opens up a new avenue for highly accurate, non-invasive speech neuroprostheses for individuals with speech disabilities by using non-invasive measurements. However, since our study did not include data from individuals with speech disabilities, this hypothesis remains to be tested. Furthermore, it is essential to validate the potential of non-invasive speech decoding using the data collection methods for attempted speech that have been employed in invasive measurements. This validation is crucial for broader applications in patients with speech disabilities.

\section{Discussion} \label{sec:discussion}
Traditionally, the development of speech neuroprosthesis has relied on invasive measurements, such as ECoG \citep{anumanchipalli2019speech,makin2020machine,moses2021neuroprosthesis,metzger2023high} and multi-unit recordings \citep{willett2021high,willett2023high}, to capture neural activity during attempted speech. However, the physical and psychological burdens associated with invasive measurements have made these technologies impractical for many patients. On the other hand, non-invasive measurements like EEG have been considered infeasible due to their low signal quality, limiting their application to classification tasks with only a few dozen classes, far from the vocabulary sizes needed for everyday conversation. This has led to the belief that developing a BCI capable of supporting the vocabulary required for everyday conversation using EEG is unfeasible. Challenging this assumption, our study collected 175 hours of EEG data during speech from the same participant, achieving 48.5\% top-1 and 76.0\% top-10 accuracy in a 512-phrase classification task. Crucially, our results suggest that decoding accuracy can continue to increase significantly with further training data, based on a scaling law between performance and data size. Moreover, we demonstrated that identifying the onset of speech---a significant challenge for traditional language-decoding BCI---can be achieved without special training or reliance on non-neural signals such as eye tracking data \citep{duan2023dewave,hollenstein2018zuco,hollenstein2019zuco}, given sufficient data. These findings suggest that realizing an open vocabulary non-invasive speech neuroprosthesis using EEG is feasible by increasing data length.

Decoding speech from EEG data obtained during attempted or overt speech has seen little progress over years due to concerns about the contamination of muscle activities. This is because EEG data during speech contain substantial electromyographic (EMG) signals, which can overshadow the neural signals related to speech. Consequently, there has been an inevitable concern that models trained on such data might decode speech relying on EMG signals rather than neural activity. Our study addressed this concern by training the model to actively ignore EMG-related signals by artificially mixing EMG signals from other trials to the training dataset. In support of this approach, the models trained on the EMG-mixed data failed to predict the speech only from the EMG signals as shown by the near-chance decoding accuracy from EMG signals alone (Table \ref{tab:emg-table}). These findings indicate that our model primarily relied on EEG signals. Our study was conducted with participants in a static seated position, minimizing the influence of non-speech-related EMG signals. However, for speech neuroprosthesis technology to be practical, it needs to function while the user is moving. Future research should collect data from participants speaking while moving and train models to handle such dynamic conditions, aiming to develop more versatile and practical models.

The state-of-the-art in EEG-based language decoding prior to our study, achieved by Défossez et al. \citep{defossez2023decoding}, involved decoding heard phrases from over 100 words with a top-10 accuracy of 25.7\% out of 190 segments for dataset \citep{Brennan2019eegdataset} and 17.7\%  out of 1842 segments for dataset \citep{Broderick2018eegdataset}. In contrast, we achieved a top-10 accuracy of 76.0\% for 512 segments when decoding phrases during speech. When the length of training data was reduced to 2.9 hours, comparable to what Défossez et al. used, our accuracy dropped to 10.5\% out of 190 segments and 1.40\% out of 1842 segments, lower than their reported accuracy. Additionally, the overlap rate of words between training and test data also dropped to a similar level as reported by Défossez et al\citep{defossez2023decoding}. (Appendix \ref{appdx:Performance}). These results suggest that the task difficulty of speech decoding may be harder than that of listening. We further discovered that speech decoding accuracy increases logarithmically with the word overlap rate (Appendix \ref{appdx:WordOverlap}). Given the limited number of commonly used vocabulary words, the required data length to learn language representation in the brain might be on the order of $10^2$ hours for both speech and listening. Future research should conduct similar analyses on reading data to demonstrate that the scaling law between data length and decoding accuracy universally applies to language decoding from the brain, regardless of whether it involves reading or speech.

\begin{ack}
We thank Masakazu Inoue and Sensho Nobe, for helpful discussions. This work was supported by the JST, Moonshot R\&D Grant Number JPMJMS2012. The authors declare no competing interests.
\end{ack}

\medskip

\bibliographystyle{nips}
{
\small

\bibliography{citation}
}

\newpage
\appendix

\section{EEG and EMG recording details} \label{sec:recording-detail}
Before electrode placement, the hair was shaved with a shaver and then cleaned with alcohol tissue. The size of the head was measured with a measuring tape to determine the location of the CZ, and eight electrode sheets coated with conductive gel (Elefix V, ZV-181E, NIHON KOHDEN, Japan) were attached at the marked positions. For the ground, the mastoid process behind the left ear was polished with Nuprep (Weaver and Company, US) and cleaned with alcohol tissue, and EMG/ECG/EKG electrode (Kendall\textsuperscript{TM}, CardinalHealth, US) was placed. 

\section{Preprocessing details}

\subsection{Formulation of adaptive filter} \label{sec:adaptive fitler}
In this study, we employed an adaptive filter implemented with Normalized Least-Mean-Square (NLMS) to remove EMG artifacts from the EEG data. The NLMS filter updates its weight matrix $w(t){\in}{\mathbb R}^{\rm ch\_emg \times ch\_eeg}$ according to equation \ref{eq:adaptive filter}, where the EEG data is represented by a vector $s(t){\in}{\mathbb R}^{\rm ch\_eeg}$ and the EMG data is represented by a vector $n(t){\in}{\mathbb R}^{\rm ch\_emg}$. In equation \ref{eq:adaptive filter} the adaptation coefficient $\eta$ was set to 0.1, and the parameter $\epsilon$, which prevents divergence caused by division, was set to the default value of 0.001. The implementation was based on padasip \citep{cejnek2022padasip} library. 

\begin{equation} \label{eq:adaptive filter}
    \begin{split}
    w(t+1) &= w(t) + \mu \frac{ n(t)\cdot(s(t)-\hat{s}(t) )^\top}{ \|n(t)\|^2 + \epsilon} \\
    \hat{s}(t) &= w(t)^\top \cdot n(t)
    \end{split}
\end{equation}

\section{Computational resources} \label{sec:compute resource}
All models were trained in parallel with Distributed Data Parallel (DDP) using four NVIDIA A100 GPUs (4 $\times$ 80 GB). It took approximately 40 hours to train one EEG Encoder or a diffusion vocoder with the full training dataset.
\section{EMG analysis} \label{sec:emg-test-proceduer}
Details of the sensitivity test for EMG, as described in Section \ref{sec:emg-test}. In training phase, the EEG encoder receives input $X$, defined as in Equation\ref{eq:emg}. 
\begin{equation}
\label{eq:emg}
    X = (1-\alpha) \times EEG + \alpha \times EMG
\end{equation} 
Here, the EMG is from a different segment than the EEG, and $\alpha$ is a value randomly sampled from range [0, 0.95]. In inference phase, either EEG or EMG is fed into the EEG encoder, and the classification accuracy is compared (Table \ref{tab:emg-table}) between the cases where EEG is used as input and where EMG is used as input. If the inference accuracy is higher when EMG is used compared to when EEG is used, it indicates that the learning of EEG encoder largely rely on EMG. However, if the inference accuracy is higher when EEG is used, it suggests that the learning of the EEG encoder prioritized EEG and tends to ignore EMG.
\begin{figure} [h]
        \centering
        \includegraphics[width=1\linewidth]{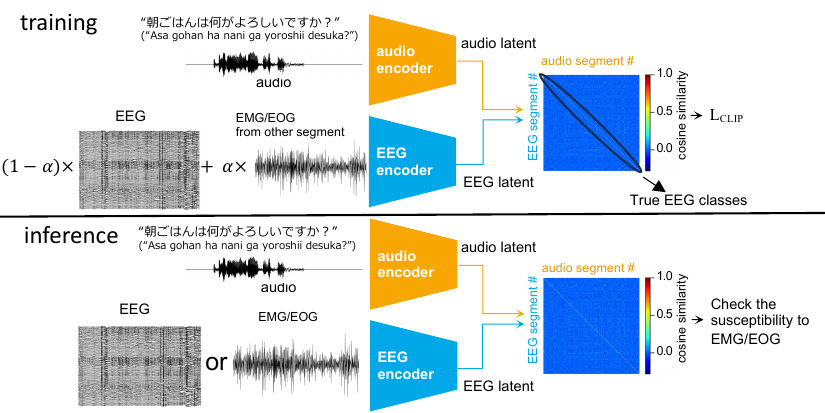}
        \caption{A schematic diagram illustrating the test procedure for evaluating the influence of EMG on the decoding process.}
       
\end{figure}

\section{Word overlap scaling}\label{appdx:WordOverlap}

\begin{figure} [h]
        \centering
        \includegraphics[width=1.0\linewidth]{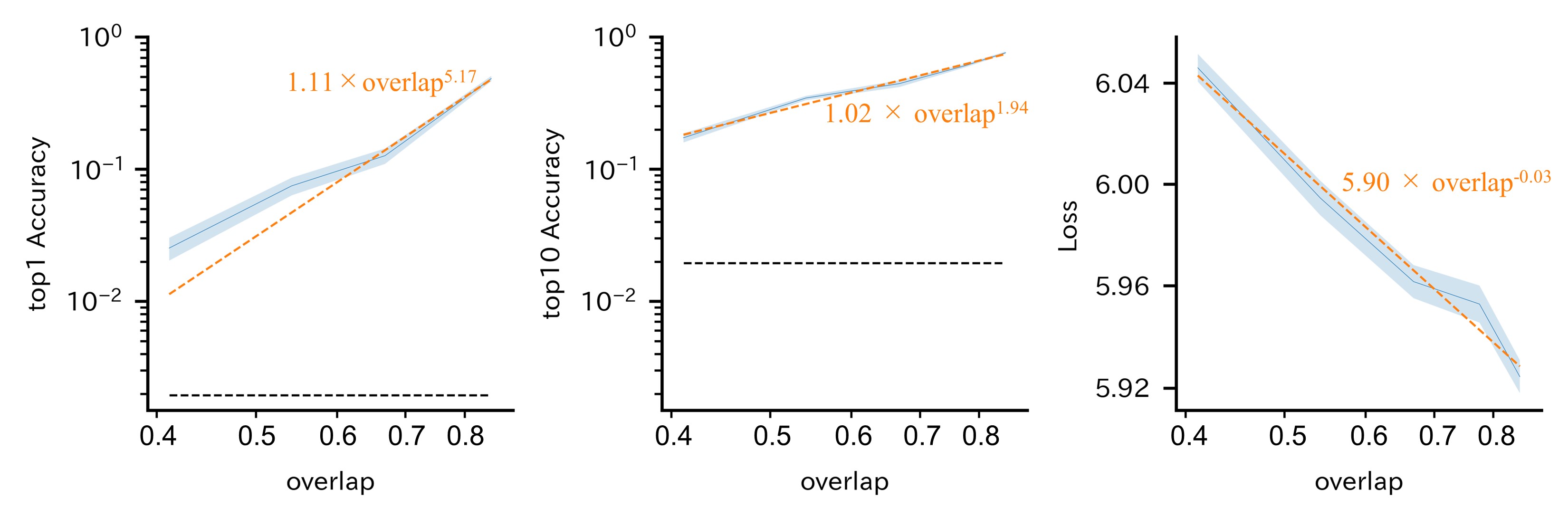}
        \caption{Word overlap increases logarithmically with recording duration.}
        \label{fig:word-overlap}
\end{figure}
We investigated the impact of word overlap between the training data and the test data on decoding accuracy. Figure \ref{fig:word-overlap} shows that the accuracy for both top-1 and top-10 increases logarithmically as the overlap increases. Correspondingly, the loss also decreases logarithmically.
\newpage
\section{Performance comparison under the same test conditions as the previous study}\label{appdx:Performance}

\begin{table} [h]
    \caption{Zero-shot segment classification performance at the same amount of training data and number of segments as previous study \citep{defossez2023decoding}.}
    \label{tab:comparison-Meta}
    \centering
    \begin{tabular}{llll}
      \toprule
      dataset & dataset size [h] & test segments & top10 (\%) \\
      \midrule
      ours (1/32) & 2.89 & 1842 & 3.64                        \\
      \citep{Brennan2019eegdataset} & 1.01 & 1842 & 17.7      \\
      ours (1/32) & 2.89 & 190 & 23.9                         \\
      \citep{Broderick2018eegdataset} & 2.20 & 190 & 25.7     \\
      \bottomrule
    \end{tabular}    
\end{table}

We compared the decoding accuracy using the same amount of data as the previous study \citep{defossez2023decoding}. Table \ref{tab:comparison-Meta} shows the comparison for two datasets \citep{Broderick2018eegdataset,Brennan2019eegdataset}. While their study focuses on decoding from hearing task, our task focuses on speech decoding.

\section{Data \& code availability}
The preprocessed EEG \& EMG, audio latent, audio waveform, and transcription used in the analysis are split into train, validation and test sets, and have been uploaded to the following data repository.
\href{https://dataverse.harvard.edu/privateurl.xhtml?token=efad3304-6bcc-4cac-a542-27078dbd09e7}{https://dataverse.harvard.edu/privateurl.xhtml?token=efad3304-6bcc-4cac-a542-27078dbd09e7}

All codes used for the analysis, including EEG and audio pre-processing, training of EEG encoder and diffusion vocoder, and inference, evaluation, and visualization, will be made available in a public repository.

\end{document}